\documentstyle[twoside,fleqn,espcrc2,epsf,psfig]{article}
\title{The Role of Splayed Disorder and Channel Flow on the Dynamics of Driven 3D Vortices}

\author{C. M. Palmer \address{Dept. of Physics, The George Washington University, Washington, DC 20052, USA}
\thanks{Present address: Dept. of Physics, University of Waterloo, Waterloo, ON N2L 3G1, Canada}
and T. P. Devereaux \address{Dept. of Physics, University of Waterloo, Waterloo, ON N2L 3G1, Canada}}
\begin{document}
\begin{abstract}
We present the results of three-dimensional molecular dynamics simulations of vortices
which indicate that, for $B$ larger than the matching field, 
the enhanced pinning effectiveness of splayed
columnar defects relative to vertical columnar defects can be explained in terms of
the existence or absence of channels through which the vortices can flow without 
encountering defects. 
\end{abstract}
\maketitle
\section{Introduction}
In recent years, experiments have shown that the critical currents of high-temperature
superconductors can be greatly enhanced by the introduction of long, straight defects
into the material, particularly if these defects are splayed at some small
angle with respect to the crystalline c-axis\cite{Kwok}.  These experiments have generally been
performed such that the magnetic field is less than the matching field, although recent efforts
have explored the opposite case\cite{Shim,Silhanek}. Explanations of this effect
are incomplete.  It has been suggested that random misalignment of the pins inhibits large-scale,
low-energy excitations, and that entanglement further prohibits vortex motion\cite{Hwa}.
At the same time tilted columns yield an increase in the bending energy of the vortices and promote hopping
at defect crossings.  Our results indicate that the physics of splay enhancement may be missing an important
ingredient.  We find that splaying columnar defects slightly
closes off channels through which vortices can flow, producing enhanced critical currents.

\section{Model}

\begin{figure}
\centerline{
\psfig{file=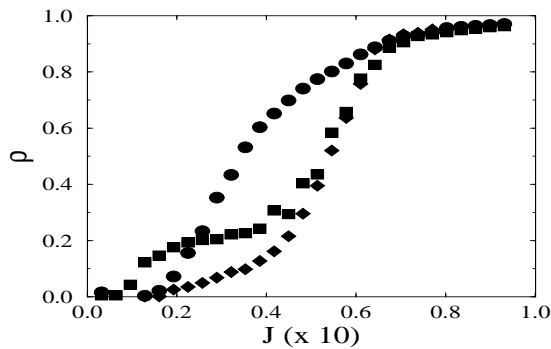,height=5.cm,width=8.cm,angle=0}}
\vskip -1.0cm
\caption[]{Resistivity ${\rho}$ as a function of driving current density J for point(circles), column(squares), and 3.6$^\circ$ splay (diamonds).}
\vskip -0.5cm
\label{fig1}
\end{figure}

We have conducted molecular dynamics simulations of driven, interacting vortices in three spatial dimensions.
The forces on the vortices result from the vortex-vortex interaction, the vortex-defect
interaction (modeled as a short-range attractive potential), elastic bending, thermal Langevin forces,
damping, and an external Lorentz force.  The energy scales are set by $V_{vortex}=(\Phi_{0}/4\pi \lambda)^{2} $,
with $\Phi_{0}$ the flux quantum and $\lambda$ the magnetic penetration depth.
Periodic boundary conditions are imposed in the a-b plane so as to maintain a constant
flux density, while open boundary conditions are imposed along the c-axis.  We have chosen
to simulate 30 vortices on 20 planes in a $16\lambda\times 16\lambda $ periodic cell, with 20 defects per plane.
At this vortex density, the vortex repulsion is strong enough to
prohibit vortex entanglement.  The defects were arranged either as vertical columns, columns
splayed at an angle $\Theta$ with respect to the c-axis,
and uncorrelated point defects. The defect radius is taken to be $\lambda$, the inter-planar spacing
is 12${\AA}$, and the coherence length ${\xi}$ is  24${\AA}$.

\section{Results}

A typical result is shown in Fig. 1 for columnar, point, and splayed defects
at low temperature.  The current density is normalized by the BCS depairing current density, while the resistivity
is normalized by the Bardeen-Stephen flux-flow resistivity.  For columnar defects at very low currents, vortices are either pinned to defects or held in place
by their mutual repulsion.  As the current density increases, some of the interstitial vortices
shear off into channels of vortex flow.  Point defects leave no open channels for vortices, but the scattering of the
defects reduces their ability to pin vortices.  Splayed defects also leave no channels, and, at small angles, are able to
effectively pin entire vortices.

\begin{figure}
\centerline{
\psfig{file=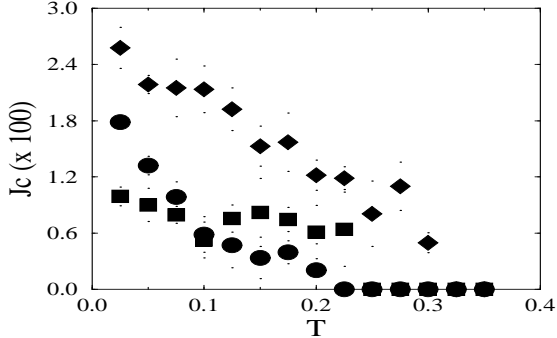,height=5.cm,width=8.cm,angle=0}}
\vskip -1.0cm
\caption[]{$J_{c}(T)$ for point (circles), column (squares), and 3.6$^\circ$ splay (diamonds).
$J_{c}$ = 0 at T = 0.225, 0.250, and 0.325 for point, column, and splay, respectively.  The error bars are roughly twice the point size.}
\vskip -0.5cm
\label{fig2}
\end{figure}

Fig. 2 shows the critical current density $J_{c}(T)$ for each configuration, where
$J_{c}$ is defined by a threshold criteria $\rho(J_{c})=0.05$.
For point defects, $J_{c}$ drops rapidly
with increasing temperature, reflecting their weak pinning efficiency.  For columnar defects,
$J_{c}$ drops slowly
with increasing temperature as more channels open for the vortices.  Splayed defects have the
largest $J_{c}$ for all temperatures.

We consider the variation in $J_{c}$ for different splay angles at low temperature in Fig. 3.
$J_{c}$
shows an enhancement at small angles due to the reduction of available channels.  For large angles,
the vortices can no longer accommodate to the defects, and splayed defects become similar
to point defects.  The value of the maximum, roughly 5$^{\circ}$, compares extremely well to the number observed in
measurements in YBa$_{2}$Cu$_{3}$O$_{7}$\cite{KrusinElbaum}.

In summary, our simulations show that the critical current enhancement from splay is the result of reduced vortex channel
flow rather than entanglement for this vortex density. We are exploring other regions of parameter space to determine
optimal conditions for splay enhancement.

\begin{figure}
\centerline{
\psfig{file=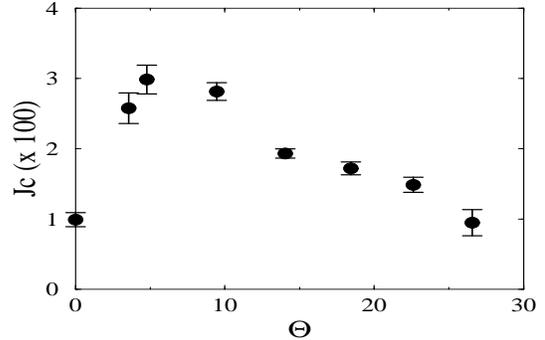,height=5.cm,width=8.cm,angle=0}}
\vskip -1.0cm
\caption[]{Critical current density as a function of splaying angle from the c-axis for T = 0.025.}
\vskip -0.5cm
\label{fig3}
\end{figure}

\end{document}